\documentclass[conference]{IEEEtran}
\IEEEoverridecommandlockouts
\usepackage{amsmath,graphicx,hyperref}
\usepackage{multirow}
\usepackage{booktabs}
\usepackage{url}
\def\BibTeX{{\rm B\kern-.05em{\sc i\kern-.025em b}\kern-.08em
    T\kern-.1667em\lower.7ex\hbox{E}\kern-.125emX}}

\title{Speaking from Coarse to Fine: Improving Neural Codec Language Model via Multi-Scale Speech Coding and Generation}

\makeatletter 
\newcommand{\linebreakand}{%
  \end{@IEEEauthorhalign}
  \hfill\mbox{}\par
  \mbox{}\hfill\begin{@IEEEauthorhalign}
}
\makeatother 

\author{
\IEEEauthorblockN{
    Haohan Guo$^\dag$\thanks{\footnotesize Work performed during the first author's internship at Xiaohongshu. This work is supported by the Centre for Perceptual and Interactive Intelligence (CPII) Ltd under the Innovation and Technology Fund.},
    Fenglong Xie$^\ddag$,
    Dongchao Yang$^\dag$,
    Xixin Wu$^\dag$,
    Helen Meng$^\dag$
}
\IEEEauthorblockA{
    $^\dag$The Chinese University of Hong Kong, Hong Kong SAR, China \\
    $^\ddag$Xiaohongshu Inc., Shanghai, China \\
    \href{mailto:hguo@se.cuhk.edu.hk}{\nolinkurl{{hguo, dcyang, wuxx, hmmeng}@se.cuhk.edu.hk}}, \href{mailto:fenglongxie@xiaohongshu.com}{{\nolinkurl{fenglongxie@xiaohongshu.com}}}
}
}

\begin{document}
%
\maketitle
\begin{abstract}

The neural codec language model (CLM) has demonstrated remarkable performance in text-to-speech (TTS) synthesis. However, troubled by ``recency bias", CLM lacks sufficient attention to coarse-grained information at a higher temporal scale, often producing unnatural or even unintelligible speech. This work proposes CoFi-Speech, a coarse-to-fine CLM-TTS approach, employing multi-scale speech coding and generation to address this issue. We train a multi-scale neural codec, CoFi-Codec, to encode speech into a multi-scale discrete representation, comprising multiple token sequences with different time resolutions. Then, we propose CoFi-LM that can generate this representation in two modes: the single-LM-based chain-of-scale generation and the multiple-LM-based stack-of-scale generation. In experiments, CoFi-Speech significantly outperforms single-scale baseline systems on naturalness and speaker similarity in zero-shot TTS. The analysis of multi-scale coding demonstrates the effectiveness of CoFi-Codec in learning multi-scale discrete speech representations while keeping high-quality speech reconstruction. The coarse-to-fine multi-scale generation, especially for the stack-of-scale approach, is also validated as a crucial approach in pursuing a high-quality neural codec language model for TTS.

\end{abstract}

\section{Introduction}

The success of large language models (LLMs) in text domain \cite{brown2020language,openai2023gpt4,touvron2023llama2} has demonstrated their great capability in discrete sequence generation. It also inspires the birth of a new text-to-speech synthesis (TTS) paradigm based on the neural codec language model (CLM) \cite{VALLEX, tortoise, lajszczak2024base}, which treats TTS as a next-token prediction task. This framework usually relies on a neural codec \cite{encodec, hifi-codec, dac} to encode the speech audio into discrete tokens, which can be incorporated with the text sequence and generated by the LM, i.e. an auto-regressive decoder. Finally, we obtain the speech audio from these generated speech tokens via the codec decoder.

However, different from the text, the discrete speech sequence is much longer to keep sufficient capacity to preserve phonetic and acoustic information. This long sequence length not only increases the complexity of TTS modeling but aggregates the ``recency bias'' of LMs \cite{peysakhovich2023attention, wang2024eliminating}, i.e. overly focusing on recent tokens during auto-regressive generation. This issue makes LM focus less on coarse-grained information \cite{guo2023msmc}, e.g. phonetics, prosody, and speaking style at higher and different temporal scales, hence causing unstable TTS performance, producing unnatural or even unintelligible speech. Although monotonic attention constraints \cite{han2024vall, du2024vall, wang2024attention} are proposed to fix stability issues, they still cannot solve ``recency bias'' fundamentally. Some works \cite{tortoise, socodec, li2024single} turn to directly model shorter speech sequences with a larger frameshift to avoid this issue, but limits the fine-grained expression of LMs in TTS. This dilemma implies the necessity of applying guidance to LMs to pay sufficient attention to both coarse-grained and fine-grained information of speech. 

In this work, we propose a novel CLM-based zero-shot TTS approach, CoFi-Speech, that generates speech in a coarse-to-fine manner via a multi-scale speech coding and generation approach. In this framework, the multi-scale speech codec, CoFi-Codec, decomposes speech into multiple discrete sequences with different temporal resolutions and decodes them back with a high reconstruction quality. Then, we propose two LM-based approaches to predict this multi-scale speech representation from coarse to fine: single-LM-based chain-of-scale generation and multiple-LM-based stack-of-scale generation. In experiments, we present subjective and objective evaluations to demonstrate that CoFi-Speech significantly outperforms baseline systems based on single-scale speech sequences on naturalness and similarity, where stack-of-scale generation performs best. Finally, we conduct detailed ablation studies to analyze multi-scale coding and generation, to further validate the effectiveness of ``speaking from coarse to fine'' in achieving high-quality CLM-based TTS.

\begin{figure*}[htp]
    \centering
    \includegraphics[width=1.0\linewidth]{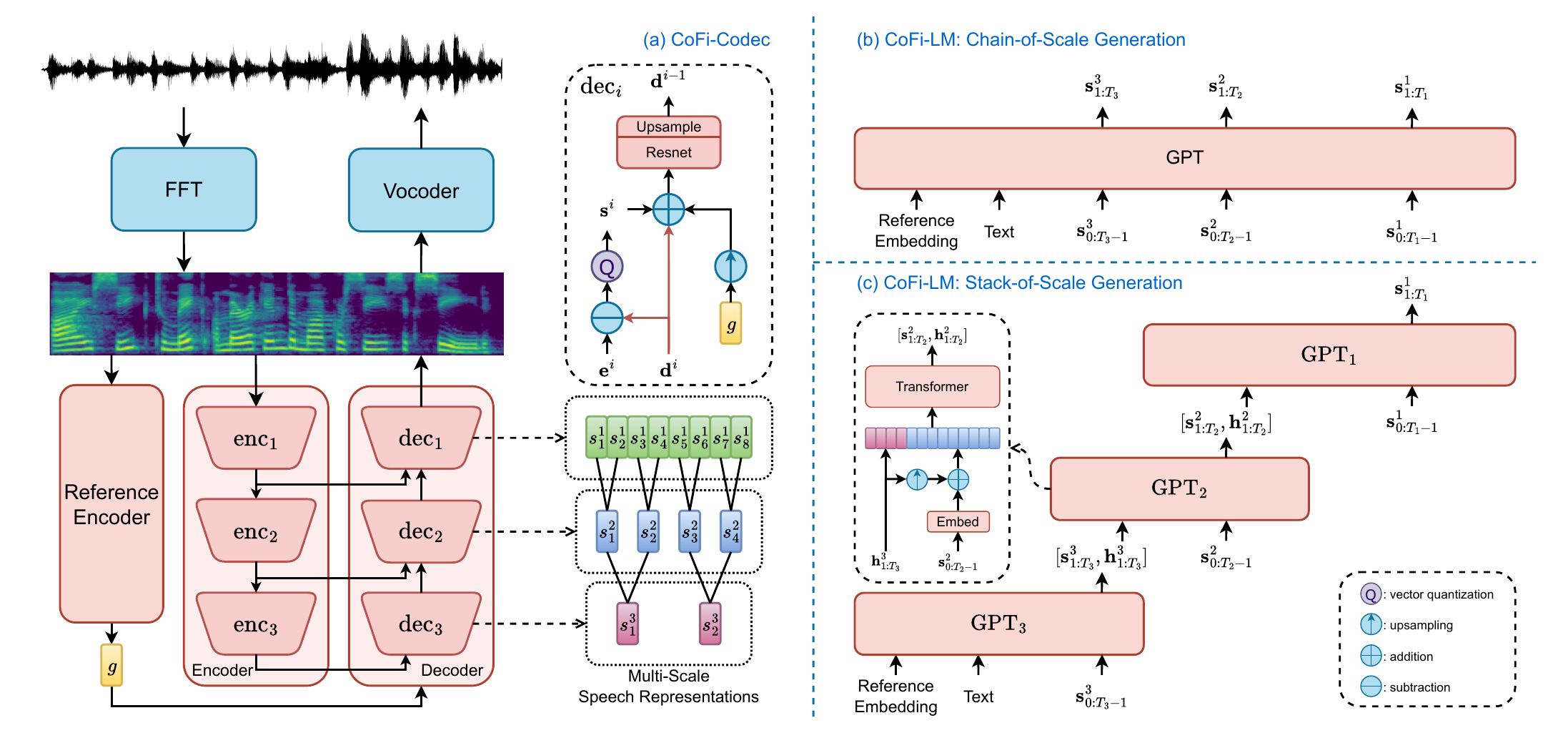}
    \caption{The system architecture of a three-scale CoFi-Speech, comprising: a CoFi-Codec for speech encoding and decoding, and two types of CoFi-LM for text-to-speech generation. The trainable modules are colored red, and each VQ operation comprises a trainable codebook.}
    \label{fig:cofi}
\end{figure*}

\section{CoFi-Speech}

\subsection{CoFi-Codec}

CoFi-Codec aims to decompose speech into multiple discrete sequences with different resolutions to provide a multi-scale speech representation. Fig. \ref{fig:cofi} (a) presents a three-scale CoFi-Codec model architecture. We first convert speech signals into the Mel spectrogram, and employ multiple encoder blocks to down-sample it into sequences at different temporal scales. Specifically, each encoder block is applied with a 1-D convolutional ResNet block and a strided convolutional layer for down-sampling. Meanwhile, we follow SoCodec \cite{socodec} to apply an ECAPA-TDNN-based \cite{dawalatabad2021ecapa} reference encoder to extract a global embedding $g$ from the Mel spectrogram to capture time-invariant information, e.g., speaker identity, global speaking style, acoustic environment, etc.

In decoding, we apply an equal number of decoder modules to quantize and decode these encoding sequences in a reversed order to reconstruct the Mel spectrogram from coarse to fine. For example, in $i$-th decoder block, we first extract the residual sequence at this stage by performing $\mathbf{e}^i - \mathbf{d}^i$, i.e. removing high-scale information from the decoding sequence $\mathbf{d}^i$ out of the encoding sequence $\mathbf{e}^i$. We then apply single-stream \cite{vqvae} or multi-stream vector quantization, e.g. RQ \cite{chen2010approximate} and PQ \cite{jegou2010product}, to obtain the quantized speech sequence $\mathbf{s}^i$. It is added with $\mathbf{d}^i$ and global embedding $g$, subsequently processed by a ResNet block and a transposed convolutional layer for up-sampling to produce the next decoding sequence $\mathbf{d}^{i-1}$. Notably, the highest-scale decoder block adopts an all-zero sequence as the decoding sequence. Finally, the Mel spectrogram is reconstructed from the decoder, and converted to the waveform via a pre-trained neural vocoder.


In training, the model tends to overfit low-scale sequences with richer information, leading to high-scale representation collapse, i.e. nothing is preserved in the sequence. Hence, we propose scale-wise nested dropout (SWND), which masks quantized sequences $\mathbf{s}_{1:b}$ given an index $b$ randomly sampled from $[0, N_s - 1]$, where $b=0$ indicates that no sequence is masked. This approach forces high-scale sequence to preserve effective information for reconstruction, avoiding representation collapse. Moreover, we apply GAN training \cite{li2024single, socodec} on CoFi-Codec to improve the generation quality of the Mel spectrogram. The loss function is written as follows:
\begin{align}
    \mathcal{L}_c &= \lambda_{vq} * \mathcal{L}_{vq} + \lambda_{reg} *\mathcal{L}_{reg} + 
    \lambda_{adv} * \mathcal{L}_{adv}
\end{align}
where $\mathcal{L}_{vq}$ is the vector-quantization loss, i.e. the averaged L2 loss between embeddings before and after VQ, $\mathcal{L}_{reg}$ is the L2-based regression loss on Mel spectrogram, and $\mathcal{L}_{adv}$ is the adversarial loss following \cite{socodec}.

\subsection{CoFi-LM}

To predict the multi-scale discrete representation in a coarse-to-fine manner, we propose CoFi-LM in two styles: chain-of-scale (CoS) generation and stack-of-scale (SoS) generation.


\noindent \textbf{Chain-of-scale generation} is a simple extension of CLM to the multi-scale speech representation, which employs one LM to predict speech sequences in descending order of the temporal scale, as shown in Fig. \ref{fig:cofi} (b). In training, the text and speech sequences are embedded into the same space and processed by a decoder-only Transformer to estimate the probabilities of the next speech tokens. Notably, for multi-stream speech sequence, we follow the ``delay pattern''\cite{copet2024simple, dang2024livespeech, lyth2024natural} to achieve the multi-stream prediction. This approach explicitly constrains the LM to model multi-scale information to alleviate ``recency bias''. However, concatenating multiple sequences leads to a longer chain, causing higher computing costs and modeling complexity. To mitigate this issue, we propose stack-of-scale generation.

\noindent \textbf{Stack-of-scale generation} employs a stack of LMs to generate speech sequences at different scales in cascade, as shown in Fig. \ref{fig:cofi} (c). It first generates the highest-scale sequence from the text and the reference embedding. The hidden states $\textbf{h}^3_{1:T_3}$ in the last transformer layer are employed to generate the next-scale sequence. It is placed at the head of the input sequence as the prompt, and also upsampled to be added to the lower-scale speech sequence to emphasize the alignment between these two sequences. In this way, we recursively generate all speech sequences from coarse to fine. This stage-wise approach enables us to better model multi-scale information while avoiding introducing long context to transformers.

Finally, we apply in-context learning (ICL) \cite{anastassiou2024seed, du2024cosyvoice} to CoFi-LMs for zero-shot TTS. Specifically, we extract speaker embedding and speech tokens from the reference audio as speech prompts. The transcript of the reference audio is also concatenated with the input text as the text prompt.

\section{Experimental Protocol}

\subsection{Training Configuration}
\label{ssec:conf}

We conduct experiments with WenetSpeech4TTS (Basic) \cite{ma2024wenetspeech4tts}, a Chinese dataset with 7k hours of speech data. All audio files are normalized to the sample rate of 16kHz and converted to 80-dim Mel spectrograms with a frameshift of 10ms. We create a byte-pair encoding (BPE) based text dictionary with 8192 tokens to encode the text.

CoFi-Codec is applied with 512-dim ResNet modules, each consisting of four residual units with two 1-D convolutional layers. $\text{enc}_1$ and $\text{dec}_1$ are applied with four more residual units to better encode and reconstruct Mel spectrograms. We train a three-scale CoFi-Codec with ordered product quantization (OPQ) \cite{socodec} to compress the Mel spectrogram into two single-stream sequences with frameshifts of 120ms and 40ms and one four-stream sequence with a frameshift of 20ms. Each stream in each sequence has a codebook with 16,384 codewords. In training, we set $\lambda_{vq}=1, \lambda_{reg}=1, \lambda_{adv}=0.1$ and use EMA \cite{vqvae} to update codebooks with the decay rate of 0.99. The discriminator in Mega-TTS \cite{jiang2023mega} is employed for adversarial training. The sampling probabilities for SWND are ${p_0 = 0.8, p_1 = 0.1, p_2 = 0.1}$. Finally, a pre-trained BigVGAN-based \cite{bigvgan} neural vocoder is employed to convert the Mel spectrogram to the waveform. 

LMs are applied with 12-layer decoder-only Transformers with a feature dimension of 1024. For multi-stream speech sequences, we apply a ``delay pattern''. We employ the sampling strategy with a top-p of 0.8, a top-k of 50, and a repetition penalty of 2.0. CoFi-Codec and LMs are trained using AdamW \cite{Loshchilov2018} for 100k iterations with a batch size of 1.6k seconds. The learning rate exponentially decays from $3\times10^{-4}$ to $1\times10^{-4}$. 

\subsection{Evaluation Metrics}

We create an 860-utterance test set with high diversity in speaking style and audio quality for system evaluation. In zero-shot TTS, each utterance is paired with another out-of-training-set audio file as the reference audio. We calculate character error rate (CER, \%), and speaker similarity (SIM, $\times10^{-2}$) to measure intelligibility, and speaker similarity.\footnote{The ASR tool for transcribing is available at \url{https://github.com/modelscope/FunASR}. The tool for extracting speaker embedding is available at \url{https://huggingface.co/Wespeaker/wespeaker-cnceleb-resnet34}} In \ref{ssec:tts_comparison}, we use a subset with 100 utterances for subjective evaluation. There are 10 native speakers invited to the test to rate each audio with a score ranging from 1 to 5 in terms of naturalness (NMOS) and speaker similarity (SMOS), respectively. In codec evaluation, We adopt Mel-cepstrum distortion (MCD, dB) and CER to measure reconstruction quality.

\section{Results}

\subsection{TTS System Comparison}
\label{ssec:tts_comparison}

As shown in Table \ref{tab:mos}, we compare CoFi-Speech with baseline systems: VALL-E and SoCodec-TTS, trained with the same dataset as ours, and the SotA industrial TTS system, CosyVoice trained with over 100k hours of data. The baseline system VALL-E auto-regressively generates an eight-stream speech sequence with a short frameshift of 20ms along both time and stream axes. This model troubled by ``recency bias'' presents serious stability issues on this challenging test set, performing the worst synthesis quality. SoCodec-TTS compresses speech into a short speech sequence with a frameshift of only 120ms to achieve stable LM inference. It presents high naturalness but low speaker similarity due to the lack of fine-grained information. In contrast, CoFi-Speech leverages both coarse-grained and fine-grained information, presenting better naturalness and similarity, especially for the SoS-based approach with the highest NMOS and SMOS, even outperforming CosyVoice trained with much more data. This significant improvement underscores the efficacy of CoFi-Speech in CLM-TTS.\footnote{Samples are available at \url{https://hhguo.github.io/DemoCoFiSpeech}}

\begin{table}[htp]
\centering
\caption{The MOS test of different CLM-based TTS systems on naturalness (NMOS) and speaker similarity (SMOS).}
\begin{tabular}{cc|cc}
\toprule
\multicolumn{2}{c|}{Systems}                                & NMOS $\uparrow$         & SMOS $\uparrow$ \\ \midrule
\multicolumn{2}{c|}{CosyVoice \cite{du2024cosyvoice} }      & 4.28          & 3.70 \\ \midrule
\multicolumn{2}{c|}{VALL-E \cite{ma2024wenetspeech4tts}}    & 3.46          & 2.92 \\
\multicolumn{2}{c|}{SoCodec-TTS \cite{socodec}}             & 4.30          & 3.51 \\ \midrule
\multirow{2}{*}{CoFi-Speech} & -CoS                         & 4.12          & 3.75 \\ 
                             & -SoS                         & \textbf{4.42} & \textbf{3.90} \\
\bottomrule
\end{tabular}
\label{tab:mos}
\end{table}

\subsection{Multi-Scale Speech Coding}
\label{ssec:msc}

\begin{figure}[htp]
    \centering
    \includegraphics[width=1.0\linewidth]{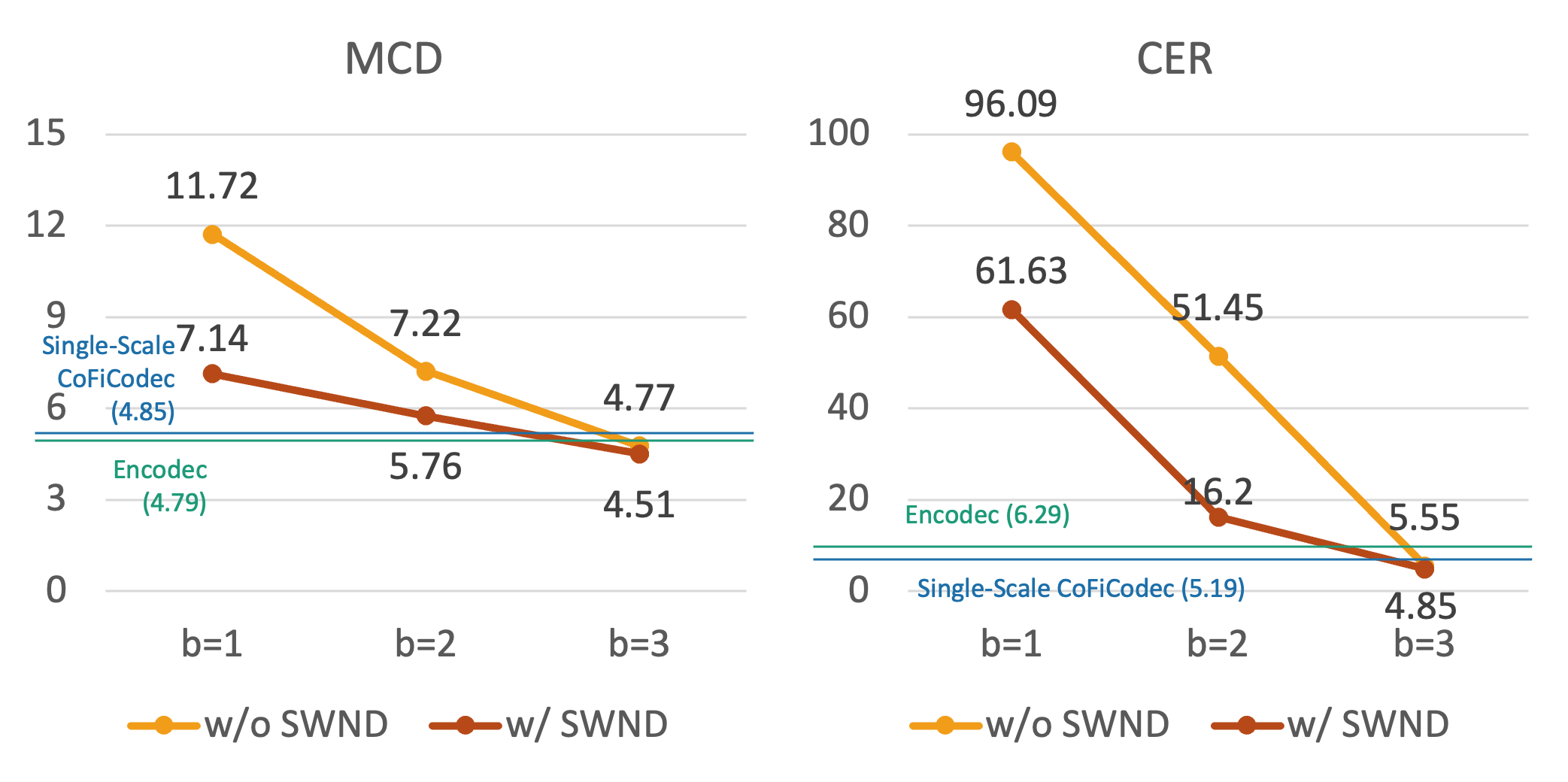}
    \caption{The objective evaluation of multi-scale coding with or without scale-wise nested dropout (SWND).}
    \label{fig:msc}
\end{figure}

We conduct objective evaluations to investigate if CoFi-Codec can decompose speech into sequences at multiple temporal scales while keeping high reconstruction quality. As shown in Fig. \ref{fig:msc}, we evaluate speech reconstructed from top-$b$ scales. For example, for a three-scale CoFi-Codec, ``$b=2$'' means that we only reconstruct speech using $\textbf{s}^3$ and $\textbf{s}^2$, and mask $\textbf{s}^1$ with all-zero vectors. The result demonstrates that CoFi-Codec learns a good multi-scale representation with the help of SWND. The high-scale sequence preserves more effective information, presenting lower reconstruction loss over that of codec without SWND. Moreover, compared with the single-scale CoFi-Codec and Encodec \cite{encodec} with a four-stream sequence on a frameshift of 20ms, CoFi-Codec presents the lowest MCD and CER. This result validates CoFi-Codec as the expected codec, providing multi-scale discrete speech representation and keeping high reconstruction quality.

\subsection{Multi-Scale Speech Generation}

\begin{figure}[htp]
    \centering
    \includegraphics[width=1.0\linewidth]{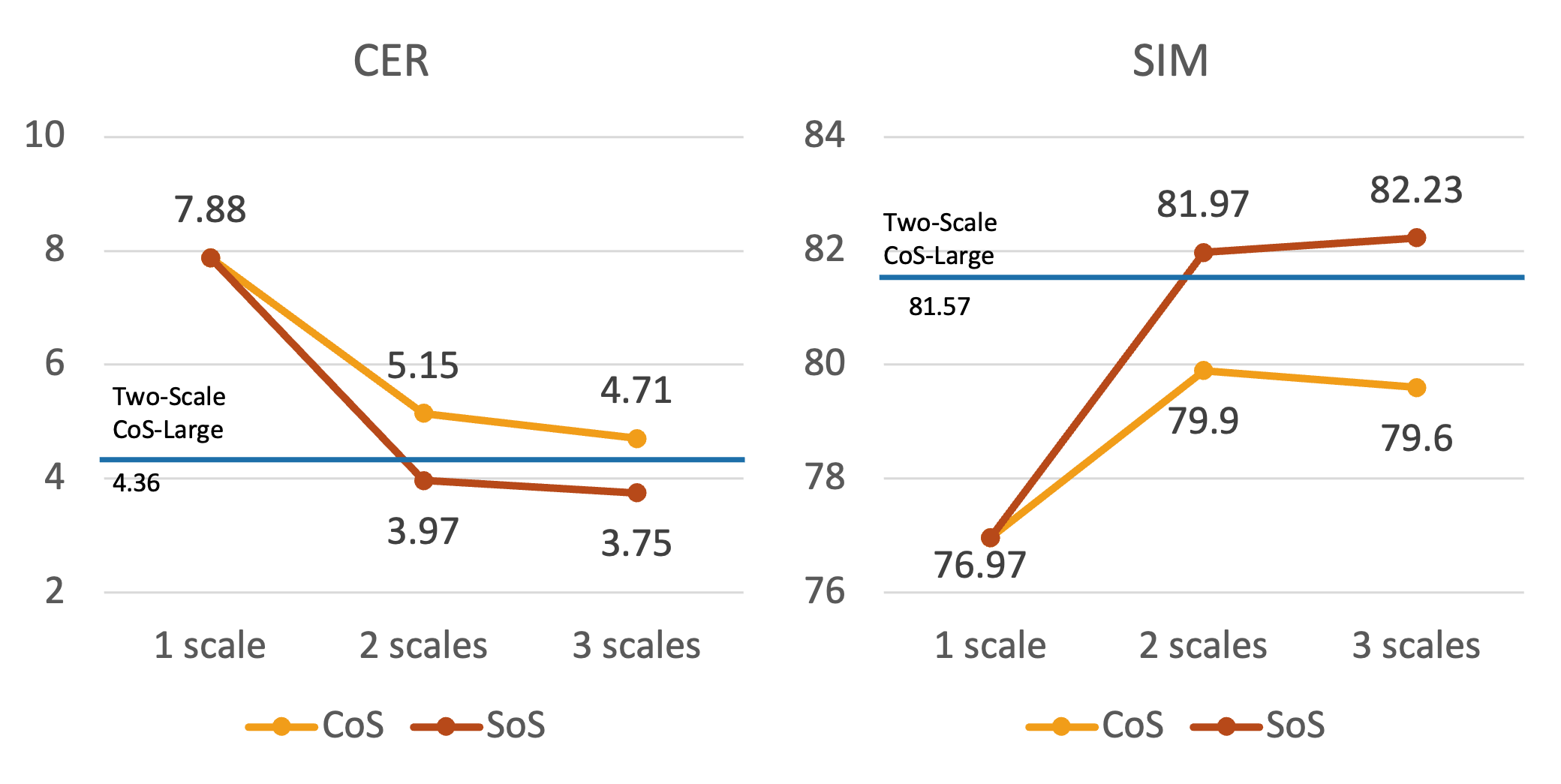}
    \caption{The objective evaluation of multi-scale generation.}
    \label{fig:msg}
\end{figure}

We also conduct objective evaluations to investigate if multi-scale speech generation can improve TTS quality. As shown in Fig. \ref{fig:msg}, we compare LMs based on three codecs: a single-scale codec with a frameshift of 20ms; a two-scale CoFi-Codec with frameshifts of 120ms and 20ms; the three-scale CoFi-Codec introduced in Sec. \ref{ssec:conf}. All codecs share the same model and training configurations. The results demonstrate that multi-scale speech generation can significantly improve TTS quality, presenting improved intelligibility and speaker similarity in CoS- and SoS-based LMs as more scales are used. Moreover, compared with CoS-LM, SoS-LM performs better with a lower CER and a higher SIM. To eliminate the impact of different numbers of parameters between CoS-LM and SoS-LM, we train a two-scale CoS-large with 24 transformer layers. It doubles the inference cost, presenting a better CER and SIM than the smaller one, but still has a gap to the two-scale and three-scale SoS-LMs. This result substantially validates the SoS as a better multi-scale generation approach.

\subsection{Recency Bias: Attention Visualization}

\begin{figure}[htp]
    \centering
    \includegraphics[width=\linewidth]{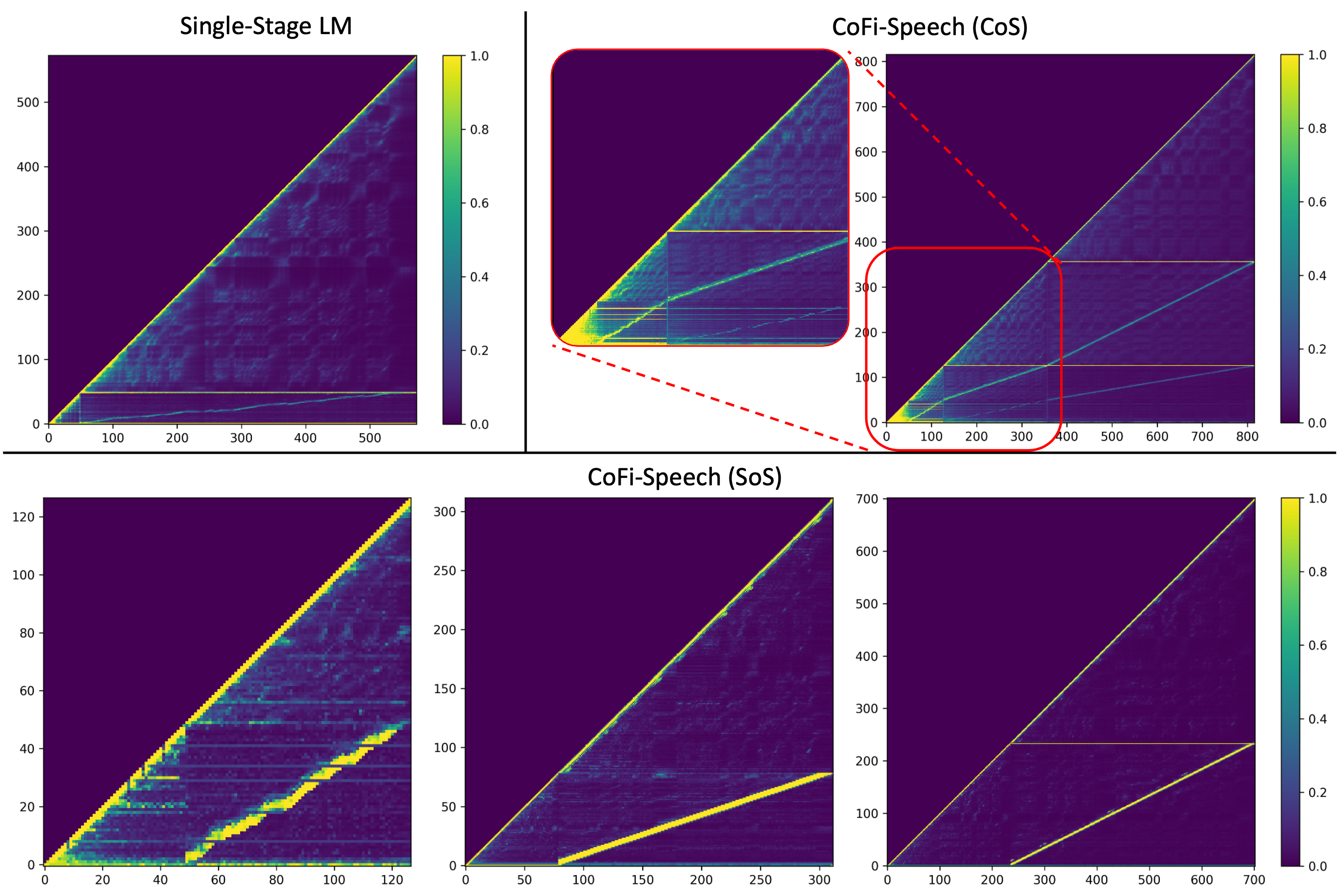}
    \caption{The aggregated attention maps of different LMs. For CoFi-Speech (SoS), maps displayed from left to right refer to $\text{GPT}_3$, $\text{GPT}_2$, and $\text{GPT}_1$.}
    \label{fig:attn}
\end{figure}

Finally, we attempt to investigate the ``recency bias'' issue in CLM-TTS by visualizing attention maps of different LMs when synthesizing the same utterance. We aggregate attention maps across all heads and layers in one model by adding them together and apply the map with a scaling factor of 0.25 and a max clipping of 1.0 for clearer visualization. As shown in Fig. \ref{fig:attn}, single-scale LM shows significant recency bias, where each speech frame pays much more attention to the nearest speech tokens and text tokens only relevant to the current pronunciation. However, in CoFi-Speech-CoS, the high-scale sequence generation presents wider attention to the text and speech context, effectively modeling coarse-grained information. It then guides the next-scale sequence generation to render fine-grained details. Similarly, CoFi-Speech-SoS employs multiple LMs to achieve coarse-to-fine generation. The first LM absorbs richer contextual information to generate the high-scale sequence and guides the next LMs to generate lower-scale sequences recursively. This validates the effectiveness of ``speaking from coarse to fine'' in addressing ``recency bias'' via explicit multi-scale modeling.

\section{Conclusions}


This work introduces CoFi-Speech, a novel approach to enhance CLM-based TTS through coarse-to-fine generation. The framework comprises two key components: 1) CoFi-Codec, a multi-scale speech codec that encodes speech into the discrete, multi-scale representation and decodes them back, and 2) CoFi-LM, which predicts this representation using two strategies: single-LM-based chain-of-scale generation and multiple-LM-based stack-of-scale generation. The zero-shot TTS experiment demonstrates that CoFi-Speech, especially for the stack-of-scale approach, significantly outperforms baseline systems using single-scale speech sequences in terms of naturalness and speaker similarity. The ablation study further validates that: 1) CoFi-Codec is capable of learning multi-scale discrete representations while preserving high-quality speech reconstruction, and 2) CoFi-LM effectively improves TTS performance via explicit multi-scale modeling. Finally, the attention visualization substantiates the efficacy of our approach in addressing ``recency bias'', providing compelling evidence for its effectiveness in improving CLM-based TTS.

\bibliographystyle{IEEEtran}
\bibliography{references}

\begin{thebibliography}{10}
\providecommand{\url}[1]{#1}
\csname url@samestyle\endcsname
\providecommand{\newblock}{\relax}
\providecommand{\bibinfo}[2]{#2}
\providecommand{\BIBentrySTDinterwordspacing}{\spaceskip=0pt\relax}
\providecommand{\BIBentryALTinterwordstretchfactor}{4}
\providecommand{\BIBentryALTinterwordspacing}{\spaceskip=\fontdimen2\font plus
\BIBentryALTinterwordstretchfactor\fontdimen3\font minus \fontdimen4\font\relax}
\providecommand{\BIBforeignlanguage}[2]{{%
\expandafter\ifx\csname l@#1\endcsname\relax
\typeout{** WARNING: IEEEtran.bst: No hyphenation pattern has been}%
\typeout{** loaded for the language `#1'. Using the pattern for}%
\typeout{** the default language instead.}%
\else
\language=\csname l@#1\endcsname
\fi
#2}}
\providecommand{\BIBdecl}{\relax}
\BIBdecl

\bibitem{brown2020language}
T.~Brown, B.~Mann, N.~Ryder, M.~Subbiah, J.~D. Kaplan, P.~Dhariwal, A.~Neelakantan, P.~Shyam, G.~Sastry, A.~Askell \emph{et~al.}, ``Language models are few-shot learners,'' \emph{Proc. NeurIPS}, vol.~33, pp. 1877--1901, 2020.

\bibitem{openai2023gpt4}
{OpenAI}, ``{GPT-4} technical report,'' \emph{arXiv preprint arXiv:2303.08774}, 2023.

\bibitem{touvron2023llama2}
H.~Touvron, L.~Martin, K.~Stone, P.~Albert, A.~Almahairi, Y.~Babaei, N.~Bashlykov, S.~Batra, P.~Bhargava, S.~Bhosale \emph{et~al.}, ``{Llama 2}: Open foundation and fine-tuned chat models,'' \emph{arXiv preprint arXiv:2307.09288}, 2023.

\bibitem{VALLEX}
Z.~Zhang, L.~Zhou, C.~Wang, S.~Chen, Y.~Wu, S.~Liu, Z.~Chen, Y.~Liu, H.~Wang, J.~Li, L.~He, S.~Zhao, and F.~Wei, ``Speak foreign languages with your own voice: Cross-lingual neural codec language modeling,'' \emph{arXiv preprint arXiv:2303.03926}, 2023.

\bibitem{tortoise}
J.~Betker, ``Better speech synthesis through scaling,'' \emph{arXiv preprint arXiv:2305.07243}, 2023.

\bibitem{lajszczak2024base}
M.~{\L}ajszczak, G.~C{\'a}mbara, Y.~Li, F.~Beyhan, A.~van Korlaar, F.~Yang, A.~Joly, {\'A}.~Mart{\'\i}n-Cortinas, A.~Abbas, A.~Michalski \emph{et~al.}, ``{BASE TTS}: Lessons from building a billion-parameter text-to-speech model on 100k hours of data,'' \emph{arXiv preprint arXiv:2402.08093}, 2024.

\bibitem{encodec}
A.~D{\'e}fossez, J.~Copet, G.~Synnaeve, and Y.~Adi, ``High fidelity neural audio compression,'' \emph{Transactions on Machine Learning Research}, 2023.

\bibitem{hifi-codec}
D.~Yang, S.~Liu, R.~Huang, J.~Tian, C.~Weng, and Y.~Zou, ``{HiFi-Codec}: Group-residual vector quantization for high fidelity audio codec,'' \emph{arXiv preprint arXiv:2305.02765}, 2023.

\bibitem{dac}
R.~Kumar, P.~Seetharaman, A.~Luebs, I.~Kumar, and K.~Kumar, ``High-fidelity audio compression with improved {RVQGAN},'' \emph{Proc. NeurIPS}, vol.~36, 2024.

\bibitem{peysakhovich2023attention}
A.~Peysakhovich and A.~Lerer, ``Attention sorting combats recency bias in long context language models,'' \emph{arXiv preprint arXiv:2310.01427}, 2023.

\bibitem{wang2024eliminating}
Z.~Wang, H.~Zhang, X.~Li, K.-H. Huang, C.~Han, S.~Ji, S.~M. Kakade, H.~Peng, and H.~Ji, ``Eliminating position bias of language models: A mechanistic approach,'' \emph{arXiv preprint arXiv:2407.01100}, 2024.

\bibitem{guo2023msmc}
H.~Guo, F.~Xie, X.~Wu, F.~K. Soong, and H.~Meng, ``Msmc-tts: Multi-stage multi-codebook vq-vae based neural tts,'' \emph{IEEE/ACM Transactions on Audio, Speech, and Language Processing}, 2023.

\bibitem{han2024vall}
B.~Han, L.~Zhou, S.~Liu, S.~Chen, L.~Meng, Y.~Qian, Y.~Liu, S.~Zhao, J.~Li, and F.~Wei, ``{VALL-E R}: Robust and efficient zero-shot text-to-speech synthesis via monotonic alignment,'' \emph{arXiv preprint arXiv:2406.07855}, 2024.

\bibitem{du2024vall}
C.~Du, Y.~Guo, H.~Wang, Y.~Yang, Z.~Niu, S.~Wang, H.~Zhang, X.~Chen, and K.~Yu, ``{VALL-T}: Decoder-only generative transducer for robust and decoding-controllable text-to-speech,'' \emph{arXiv preprint arXiv:2401.14321}, 2024.

\bibitem{wang2024attention}
H.~Wang, C.~Du, Y.~Guo, S.~Wang, X.~Chen, and K.~Yu, ``Attention-constrained inference for robust decoder-only text-to-speech,'' \emph{arXiv preprint arXiv:2404.19723}, 2024.

\bibitem{socodec}
H.~Guo, F.~Xie, K.~Xie, D.~Yang, D.~Guo, X.~Wu, and H.~Meng, ``{SoCodec}: A semantic-ordered multi-stream speech codec for efficient language model based text-to-speech synthesis,'' \emph{arXiv preprint arXiv:2409.00933}, 2024.

\bibitem{li2024single}
H.~Li, L.~Xue, H.~Guo, X.~Zhu, Y.~Lv, L.~Xie, Y.~Chen, H.~Yin, and Z.~Li, ``{Single-Codec}: Single-codebook speech codec towards high-performance speech generation,'' in \emph{Proc. Interspeech}, 2024, pp. 3390--3394.

\bibitem{dawalatabad2021ecapa}
B.~Desplanques, J.~Thienpondt, and K.~Demuynck, ``{ECAPA-TDNN:} emphasized channel attention, propagation and aggregation in {TDNN} based speaker verification,'' in \emph{Proc. Interspeech}.\hskip 1em plus 0.5em minus 0.4em\relax {ISCA}, 2020, pp. 3830--3834.

\bibitem{vqvae}
A.~van~den Oord, O.~Vinyals, and K.~Kavukcuoglu, ``Neural discrete representation learning,'' in \emph{Proc. NeurIPS}, 2017.

\bibitem{chen2010approximate}
Y.~Chen, T.~Guan, and C.~Wang, ``Approximate nearest neighbor search by residual vector quantization,'' \emph{Sensors}, vol.~10, no.~12, pp. 11\,259--11\,273, 2010.

\bibitem{jegou2010product}
H.~Jegou, M.~Douze, and C.~Schmid, ``Product quantization for nearest neighbor search,'' \emph{IEEE Trans. Pattern Anal. Mach. Intell.}, vol.~33, no.~1, pp. 117--128, 2010.

\bibitem{copet2024simple}
J.~Copet, F.~Kreuk, I.~Gat, T.~Remez, D.~Kant, G.~Synnaeve, Y.~Adi, and A.~D{'e}fossez, ``Simple and controllable music generation,'' \emph{Proc. NeurIPS}, vol.~36, 2024.

\bibitem{dang2024livespeech}
T.~Dang, D.~Aponte, D.~Tran, and K.~Koishida, ``{LiveSpeech}: Low-latency zero-shot text-to-speech via autoregressive modeling of audio discrete codes,'' in \emph{Proc. Interspeech}, 2024, pp. 3395--3399.

\bibitem{lyth2024natural}
D.~Lyth and S.~King, ``Natural language guidance of high-fidelity text-to-speech with synthetic annotations,'' \emph{arXiv preprint arXiv:2402.01912}, 2024.

\bibitem{anastassiou2024seed}
P.~Anastassiou, J.~Chen, J.~Chen, Y.~Chen, Z.~Chen, Z.~Chen, J.~Cong, L.~Deng, C.~Ding, L.~Gao \emph{et~al.}, ``{Seed-TTS}: A family of high-quality versatile speech generation models,'' \emph{arXiv preprint arXiv:2406.02430}, 2024.

\bibitem{du2024cosyvoice}
Z.~Du, Q.~Chen, S.~Zhang, K.~Hu, H.~Lu, Y.~Yang, H.~Hu, S.~Zheng, Y.~Gu, Z.~Ma \emph{et~al.}, ``{CosyVoice}: A scalable multilingual zero-shot text-to-speech synthesizer based on supervised semantic tokens,'' \emph{arXiv preprint arXiv:2407.05407}, 2024.

\bibitem{ma2024wenetspeech4tts}
L.~Ma, D.~Guo, K.~Song, Y.~Jiang, S.~Wang, L.~Xue, W.~Xu, H.~Zhao, B.~Zhang, and L.~Xie, ``{WenetSpeech4TTS}: A 12,800-hour {Mandarin TTS} corpus for large speech generation model benchmark,'' in \emph{Proc. Interspeech}, 2024, pp. 1840--1844.

\bibitem{jiang2023mega}
Z.~Jiang, J.~Liu, Y.~Ren, J.~He, Z.~Ye, S.~Ji, Q.~Yang, C.~Zhang, P.~Wei, C.~Wang, X.~Yin, Z.~MA, and Z.~Zhao, ``Mega-{TTS} 2: Boosting prompting mechanisms for zero-shot speech synthesis,'' in \emph{Proc. ICLR}, 2024.

\bibitem{bigvgan}
S.-g. Lee, W.~Ping, B.~Ginsburg, B.~Catanzaro, and S.~Yoon, ``{BigVGAN}: {A} universal neural vocoder with large-scale training,'' in \emph{Proc. ICLR}, 2023.

\bibitem{Loshchilov2018}
I.~Loshchilov and F.~Hutter, ``{Decoupled weight decay regularization},'' in \emph{Proc. ICLR}, 2018.

\end{thebibliography}

\end{document}